\documentclass[fleqn,usenatbib,letters]{mnras}

\usepackage{newtxtext,newtxmath}

\usepackage[T1]{fontenc}
\usepackage{ae,aecompl}

\usepackage{graphicx}	
\usepackage{amsmath}	
\usepackage{amssymb}	
\usepackage{xcolor}
\usepackage{ulem}
\usepackage{verbatim}
\newcommand{\HI}{\ion{H}{i}}

\definecolor{orange}{rgb}{1.0,0.3,0.0}

\newcommand{\cmsq}{\ensuremath{{\rm cm}^{-2}}}

\title[H$_2$ in composite DLA spectra]
{Constraining the H$_2$ column density distribution at z$\sim$3 from composite DLA spectra}

\author[S. A. Balashev and P. Noterdaeme]{S.~A.~Balashev\,$^{1}$ and P.~Noterdaeme\,$^{2}$
	\\
	$^{1}$Ioffe Institute, {Polytekhnicheskaya ul. 26}, 194021 Saint-Petersburg, Russia -- email: s.balashev@gmail.com \\
	$^{2}$Institut d'Astrophysique de Paris, CNRS-UPMC, UMR7095, 98bis bd Arago, 75014 Paris, France -- email: noterdaeme@iap.fr 
}

\date{Accepted 2018 April 11. Received 2018 March 30; in original form 2018 February 14}

\pubyear{2018}

\begin{document}
\label{firstpage}
\pagerange{\pageref{firstpage}--\pageref{lastpage}}
\maketitle

\begin{abstract}
	We present the detection of the average H$_2$ absorption signal in the overall population of neutral gas absorption systems at $z\sim 3$ using composite absorption spectra built from the Sloan Digital Sky Survey-III damped Lyman-$\alpha$ catalogue. We present a new technique to directly measure the H$_2$ column density distribution function $f_{\rm H_2}(N)$ from the average H$_2$ absorption signal.
Assuming a power-law column density distribution, we obtain a slope $\beta = -1.29 \pm 0.06(\rm stat) \pm 0.10 (\rm sys)$ and an incidence rate of strong H$_2$ absorptions (with $N$(H$_2)\gtrsim 10^{18}$~\cmsq) to be $4.0 \pm 0.5(\rm stat) \pm 1.0 (\rm sys)\,\%$ in H\,{\sc i} absorption systems with $N($H\,{\sc i})$\ge 10^{20}$~cm$^{-2}$.  
Assuming the same inflexion point where $f_{\rm H_2}(N)$ steepens as at $z=0$, we estimate that the cosmological density of H$_2$ in the column density range $\log N(\rm H_2) (\cmsq)= 18-22$ is $\sim 15\%$ of the total. We find one order of magnitude higher H$_2$ incident rate in a sub-sample of extremely strong DLAs ($\log N(\HI) (\cmsq) \ge 21.7$), which, together with the the derived shape of $f_{\rm H_2}(N)$, suggests that the typical H\,{\sc i}-H$_2$ transition column density in DLAs 
is $\log N({\rm H}) (\cmsq) \gtrsim22.3$ in agreement with theoretical expectations for the average (low) metallicity of DLAs at high-$z$.
\end{abstract}




\section{Introduction}

Recent works have emphasized the need to split the interstellar medium (ISM) into its atomic and molecular hydrogen components for a proper modelling of the formation and evolution of galaxies \citep[e.g.][]{Lagos2011}. Observational constraints on these phases are available for large samples of nearby galaxies and with sub-kpc resolution \citep[e.g.][]{Bigiel2008}. However, similar observations remain limited at high redshifts and are generally available only for a single phase at a time. 
The \HI\ gas is blindly probed at high redshifts thanks to the damped Lyman-$\alpha$ absorption systems (DLAs) in the spectra of background sources such as quasars. The {\sl global} \HI\ column density distribution and total cosmological density is now well constrained \citep[e.g.][]{Prochaska2009,Noterdaeme2012}, but this technique does not provide the \HI\ content for a given galaxy. Indeed, despite recent progress \citep[see e.g.][and references therein]{Krogager2017}, the association between DLAs and galaxies remains difficult. 

The molecular hydrogen component at high redshifts is indirectly probed using CO emission as a proxy in emission-selected galaxies. However, \HI\ maps of the same galaxies (which, due to their selection in emission, are likely to be biased towards high masses) will have to await for future facilities such as the Square Kilometre Array, and even then, the spatial resolution is much coarser than the typical scale of ISM clouds. 

While H$_2$ is very hard to detect in emission because of the lack of a dipole moment, it does imprint resonant electronic absorption bands in the UV (the so-called Lyman and Werner bands). Since these bands fall in-between the wealth of \HI\ absorption lines from the intergalactic medium (i.e. the Lyman-$\alpha$ forest), the detection of H$_2$ is eased by the use of high resolution spectroscopy \citep[e.g.][]{Levshakov1985, Ledoux2006, Balashev2017}. However, this is extremely time-consuming since the detection rate is small ($\lesssim 10 \%$), i.e. a large number of DLAs need to be surveyed to constrain H$_2$ statistics. Dedicated observations plus archival high-resolution data have been used for such purposes \citep{Petitjean2000,Ledoux2003,Noterdaeme2008} but there are inevitably selection effects that are hard to control, in particular when using archival data. 
\citet{Jorgenson2014} have presented a more homogeneous search of a mix of low and high-resolution spectra. However, the statistics are sensitive to the exact determination of detection limit, and remain limited with a single (previously-known) H$_2$ detection.

In this letter, we use a very large and homogeneous sample of damped Lyman-$\alpha$ systems to 
detect the weak mean-signature of molecular hydrogen in composite spectra and derive the detection rate of H$_2$ in DLAs and, for the first time, the H$_2$ column density distribution at high redshift.

\section{Detection of the H$_2$ signal in composite DLA spectra}

Analysing composite spectra built by averaging a large amount of individual 'poor' quality (i.e. with low S/N and intermediate resolution) spectra \citep[e.g.,][]{Nestor2003, Wild2006, Noterdaeme2010a, Rahmani2010, Joshi2017, Mas-Ribas2017} allows us to directly obtain information on the mean properties of a population of quasar absorbers without the need of time-consuming follow-up observations. A main strength of this method is that features that cannot be confidently detected in individual spectra or are present only in a fraction of them become visible in the high S/N composite spectrum. For a given species with a column density distribution $f(N)$, the resulting composite absorption profile is 
\begin{equation}
\label{stackspectrum}
S(\lambda) = \int\limits_{N_{\rm low}}^{N_{\rm up}} f(N) \int  R(\lambda-\lambda') e^{-\tau(N, \lambda')} d\lambda' dN,
\end{equation}
where $\tau(N, \lambda')$ denotes the profile of a single absorption system with column density $N$, $R$ is the instrument line spread function.
We remark that the composite profile differs from the profile of an absorption system with column density equal to the average value. 

\begin{figure}
\centering		
\includegraphics[width=\hsize]{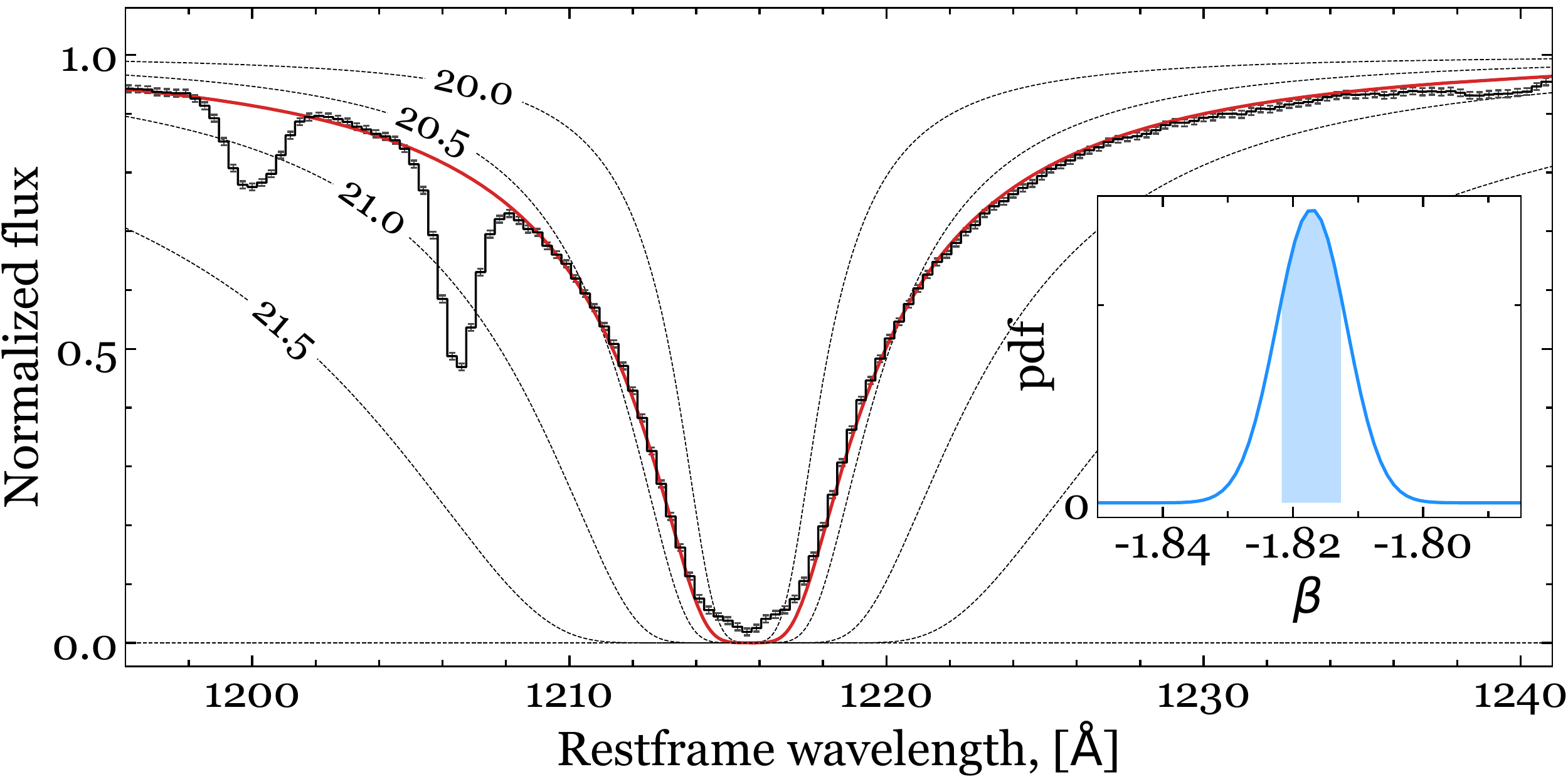}
		\caption{The fit to \HI\, Ly$\alpha$ line in the composite DLA spectrum. The red profile corresponds to the best fit using a power-law column density distribution with slope $\beta=-1.82$. 
        As discussed by \citet{Mas-Ribas2017}, the bottom of the Ly-$\alpha$ line does not go exactly to zero flux. This can be due zero-flux calibration uncertainties \citep{Paris2012}, leaking UV emission from quasar host \citep{Cai2014} and to residual contribution from false-positive DLAs. The corresponding pixels are not used to constrain the fit. For illustrative purposes, we show the Ly-$\alpha$ profiles corresponding to a single absorption system with $\log N(\HI)=20, 20.5, 21, 21.5$. 
        Inlay: the posterior probability distribution of $\beta$ estimated from the profile fitting. The shaded region corresponds to the 0.683 confidence interval. 
       }
		\label{HIstack}
\end{figure}

Before analysing H$_2$, we check whether our method allows to retrieve the \HI\ distribution function, which is known from the counting of individual systems. We use the composite SDSS-DLA spectrum from \citet{Mas-Ribas2017}, built using the spectra of  $\sim 27,000$ DLAs from \citet{Noterdaeme2012}. Details about the rejection of low S/N spectra, weighting and contribution of each spectrum at a given wavelength are given in \citet{Mas-Ribas2017}. Although this composite includes systems down to $\log N(\HI)=20$, i.e. below the conventional definition for DLAs ($\log N(\HI)\ge 20.3$), we refer to the corresponding sample as the {\sl DLA} sample for simplicity. 
We fitted the average Lyman-$\alpha$ absorption, shown in Fig.~\ref{HIstack} using a truncated power law for the column density distribution
\begin{equation}\label{powerlaw}
f(N) =  C \cdot N^{\beta},
\end{equation}
where $C$ is the proper normalization constant for a power law distribution in the fixed range $[N_{\rm low},N_{\rm up}]$.
The lower cut-off is set to $\log N_{\rm low}(\HI)=20$, i.e. the minimum column density of systems used in the stack. We set $\log N_{\rm up}(\HI)=22$ since $f(N_{\rm HI})$ is known to further steepen at higher column densities \citep{Noterdaeme2012}. 
Systems with such high column densities contribute little 
to the composite anyway and the fit is not very sensitive to this upper bound. 
We obtain a best-fit power-law slope of $\beta=-1.817\pm0.005$, where the very small formal statistical error do not take into account continuum placement and  redshift measurement uncertainties. 
The derived best-fit value agrees remarkably well
with the distribution derived from counting and fitting each absorber individually.

\begin{figure*}
\centering
\setlength{\tabcolsep}{1pt}
    \begin{tabular}{cc}
    \includegraphics[trim={0.0cm 1.3cm 0.2cm 0.2cm},clip,width=0.75\textwidth]{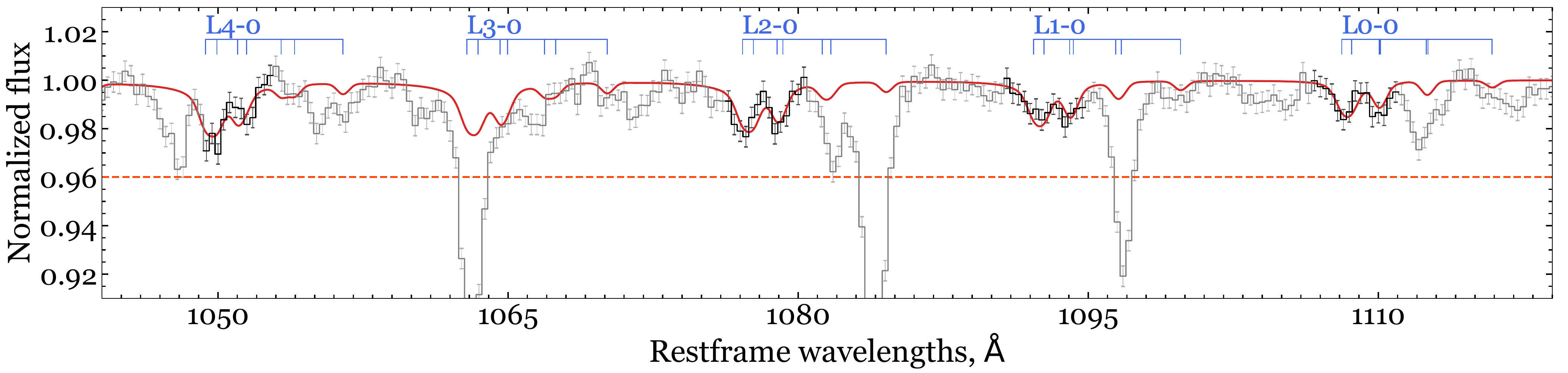}
    & \includegraphics[trim={0.5cm 1.8cm 0.5cm 0.5cm},clip,width=0.25\textwidth]{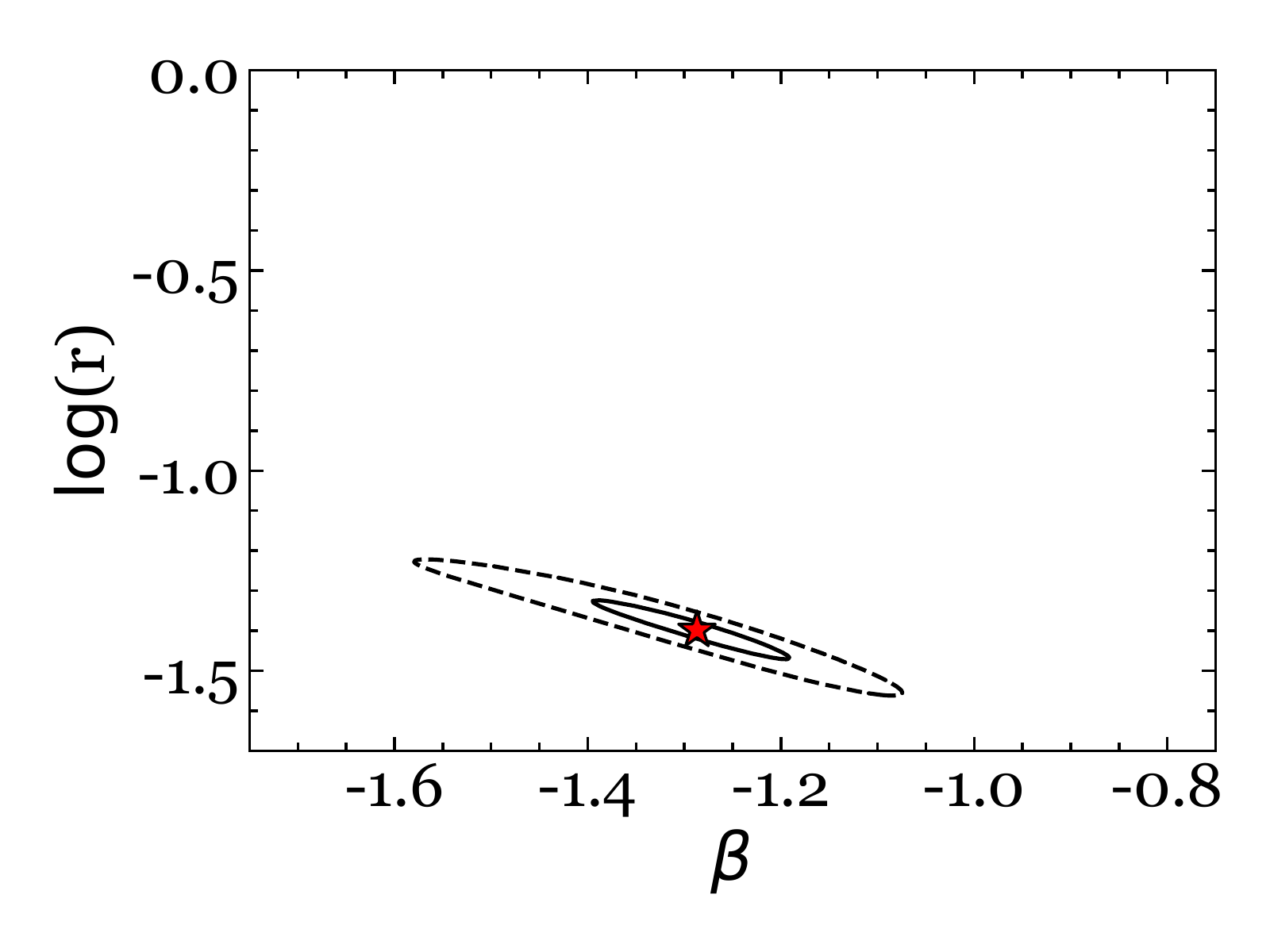}\\
     \includegraphics[trim={0.0cm 1.3cm 0.2cm 0.2cm},clip,width=0.75\textwidth]{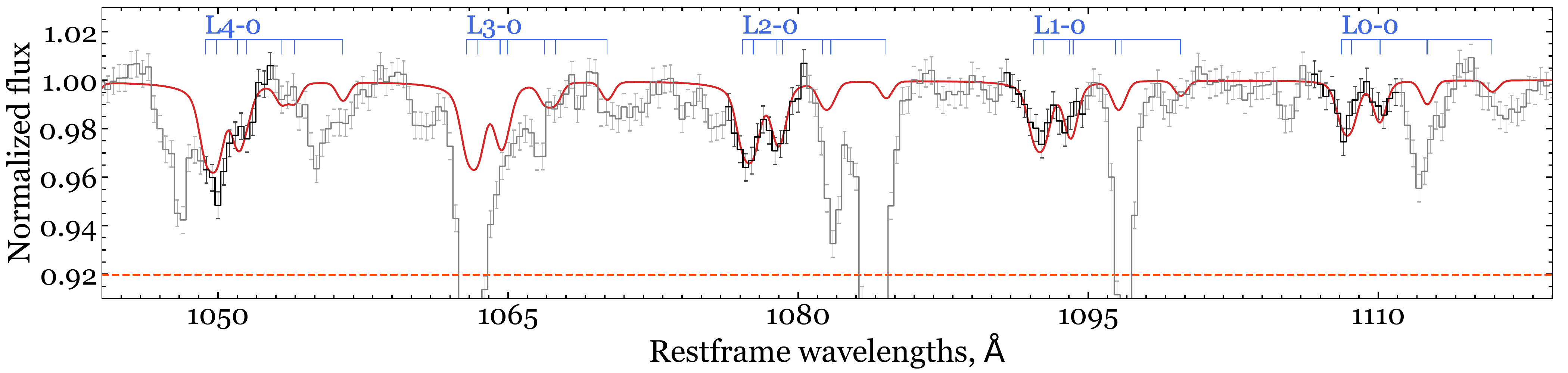}
    & \includegraphics[trim={0.5cm 1.8cm 0.5cm 0.5cm},clip,width=0.25\textwidth]{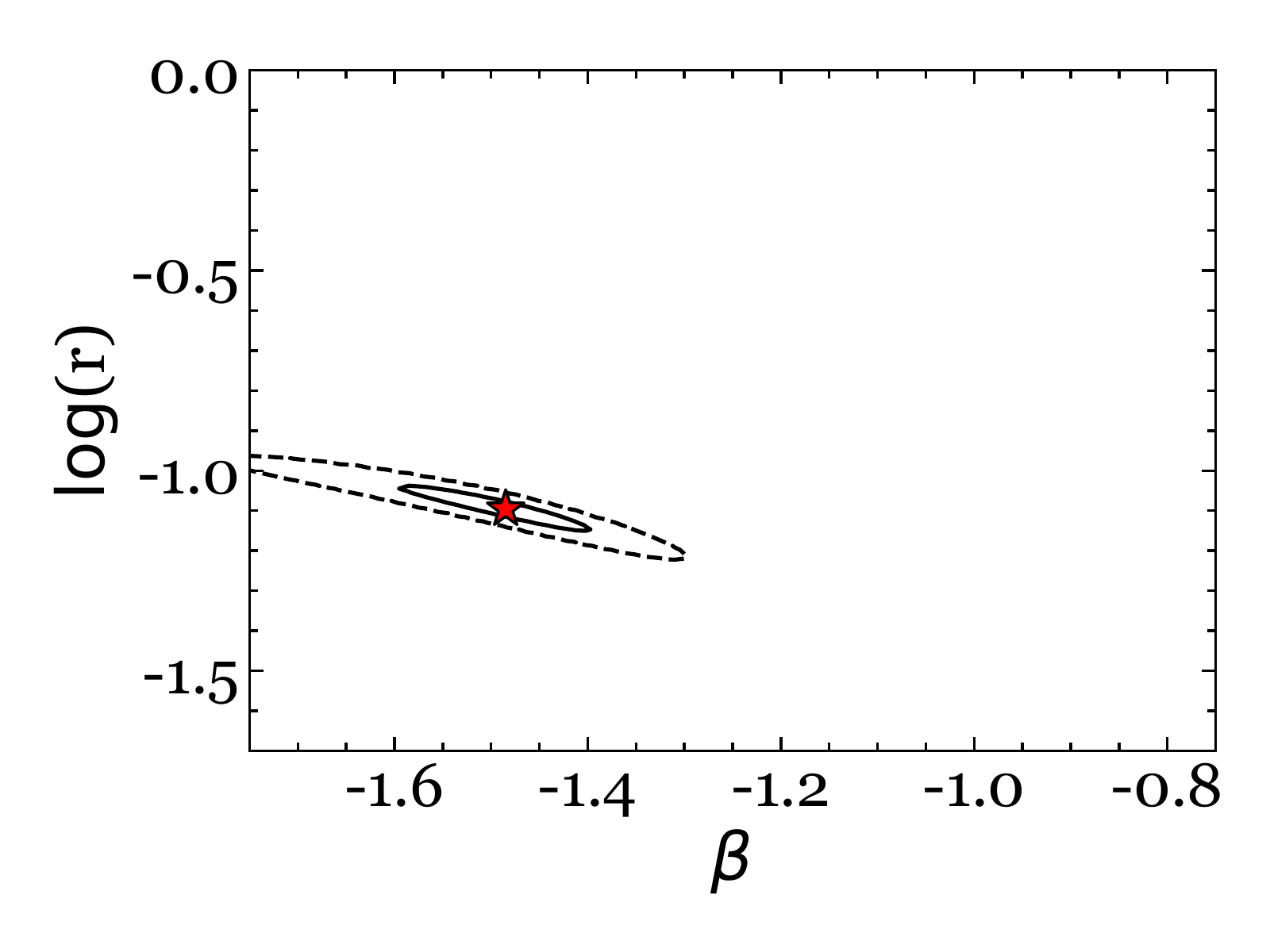}\\
    \includegraphics[trim={0.0cm 0.2cm 0.2cm 0.2cm},clip,width=0.75\textwidth]{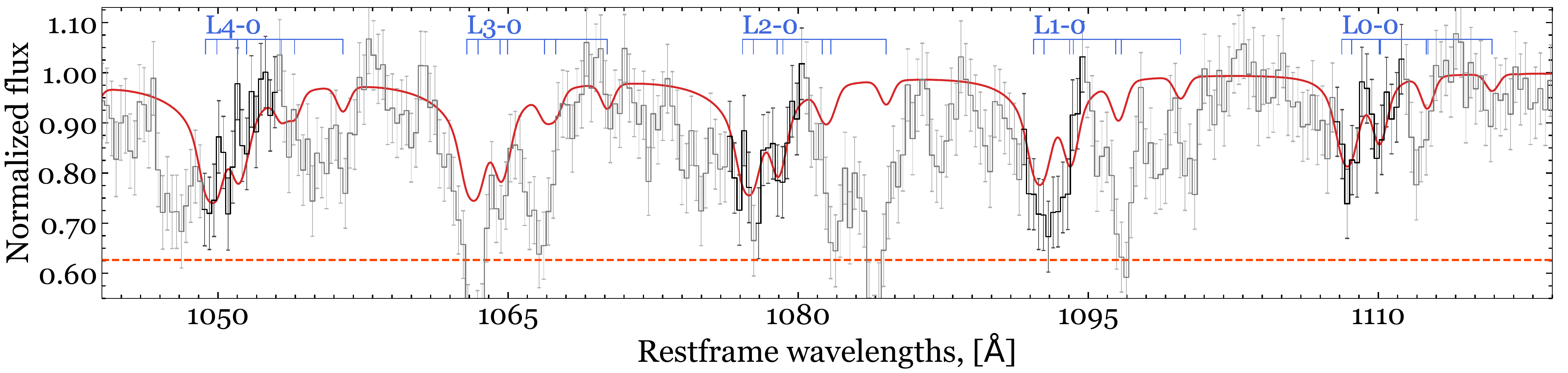}
    & \includegraphics[trim={0.5cm 0.7cm 0.5cm 0.5cm},clip,width=0.25\textwidth]{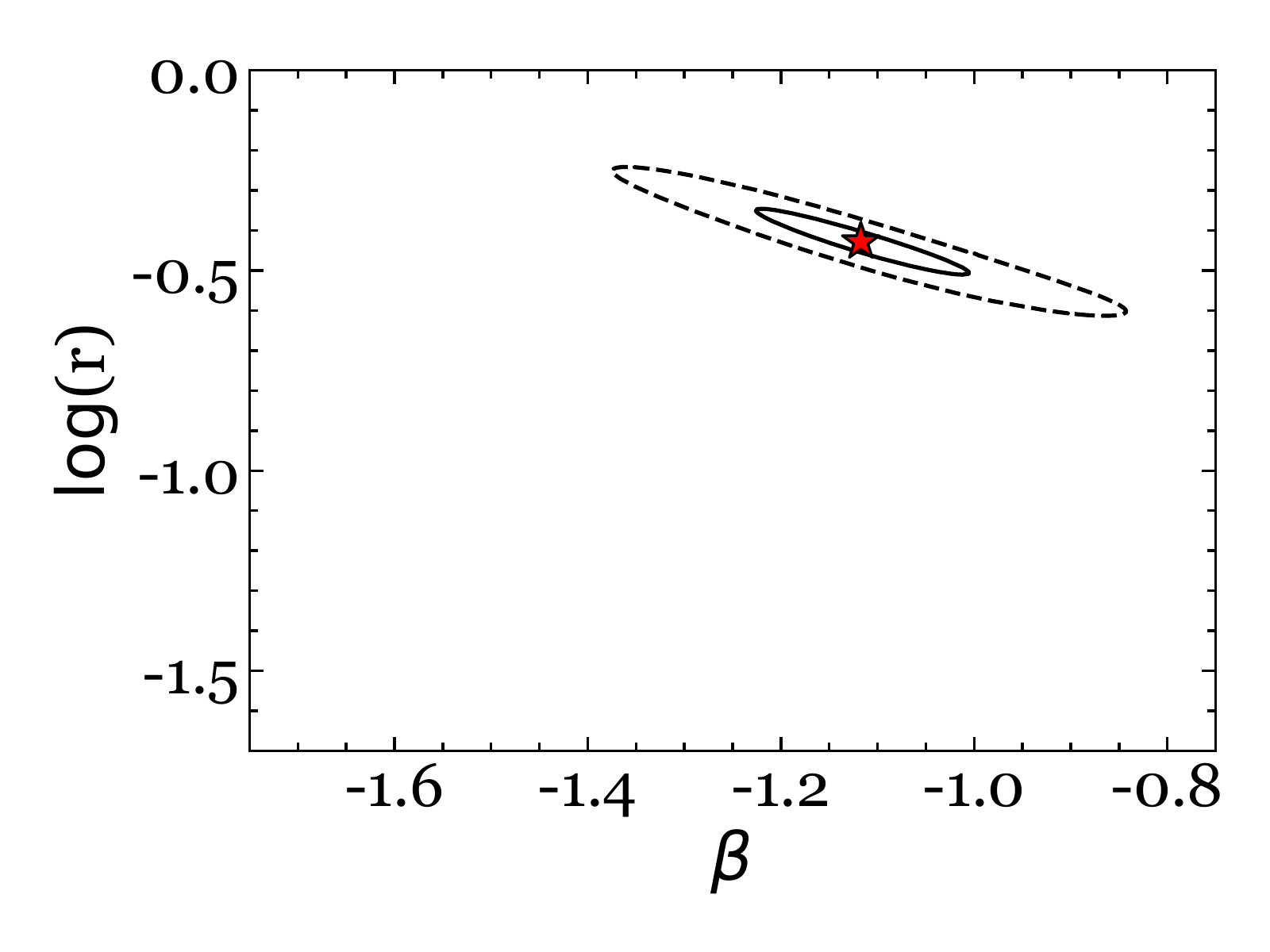}\\
        \end{tabular}
\setlength{\tabcolsep}{6pt}

\caption{Detection of H$_2$ in three composite spectra. From top to bottom: {DLAs}, {\sl metal-DLAs} and {\sl ESDLAs} samples. Note the much wider y-axis range for the {\sl ESDLA} composite.
{\sl Left panels:} black lines represent the H$_2$ J=0,1 regions that were used to constrain the fit (i.e. not blended with other absorption lines such as Fe\,{\sc ii} at 1063, 1081, 1096 \AA), the rest of the composite spectra being shown in grey. Red lines show the best-fit composite H$_2$ profiles. The orange dashed horizontal lines correspond to the best-fit value of effective non coverage factor, $1-r$. {\sl Right panels:} constraints on $r$ and $\beta$. The best-fit value is shown as a red star, and the 1\,$\sigma$ and 3\,$\sigma$ contours are shown by solid and dashed lines, respectively. \label{SDSS}}
 \end{figure*}

Next, we apply our method to H$_2$. The wavelength region where H$_2$ bands 
are covered and not blended with metal lines is shown in Fig.~\ref{SDSS}. Despite 
the weak lines, the average H$_2$ signal is clearly detected thanks to the high S/N ratio of the composite. We also show the {\sl metal-DLA} composite from \citet{Mas-Ribas2017}, that corresponds to a sub-sample of $\sim$12,000 DLA candidates in which prominent 
metal lines are detected. Finally, we created a composite spectrum of extremely strong DLAs (ESDLAs, with $\log N(\HI) (\cmsq) \ge 21.7$) visually inspected and selected from DLAs automatically discovered in the SDSS by \citet{Garnett2017} and \citet{Parks2018} (DR12) as well as by the procedure of \citet{Noterdaeme2012} applied to DR14 \citep[as in][]{Noterdaeme2014}. 
Because of the small number of systems -- only 51 ESDLAs contribute to the wavelength region of H$_2$ lines -- the composite was carefully built after visually normalising the quasar continuum using B-splines for each spectrum and calculating the composite at each wavelength using a median average. Using other averaging schemes (e.g. simple mean or $\sigma$-clipping) results in marginal differences only.
We already note that the H$_2$ signal is about twice stronger in the {\sl metal-DLA} composite and an order of magnitude stronger in the {\sl ESDLA} composite than compared to {\sl DLA} composite. 

In order to quantify the H$_2$ content of these samples ({\sl DLAs}, {\sl metal-DLAs} and {\sl ESDLAs}), we again use the truncated power-law distribution and set the upper-bound to log $N_{\rm up}($H$_2)=22$ -- several times higher that 21.3 -- the highest values observed so far in high-$z$ DLAs (\citealt{Balashev2017}, Ranjan et al., in prep).  
We note that H$_2$ column densities with values higher than this would induce complete blending of the lines of Lyman and Werner bands and hence should affect mostly the apparent continuum normalisation, which is not trivial to quantify. 
It is also expected that current optical samples will be biased against such systems since they strongly extinguish the background quasar light. 
The lower bound is set to $\log N_{\rm low}($H$_2)=18$. Due to the very steep change in the self-shielding in the range $\log N($H$_2)\sim 16-18$, the distribution of H$_2$ column densities in different environments (e.g. Solar neighbourhood, Magellanic clouds and high-$z$ DLAs) tends to be bimodal, with values either above $\log N($H$_2)\sim18$ (self-shielded regime) or well below $\log N($H$_2)\sim16$. 

High-resolution studies have also shown that H$_2$ is actually not detected in most DLAs down to detection limits of $\log$N(H$_2$)$\sim$13. We therefore introduce an incidence parameter $r$ that represents the fraction of DLAs that have $\log N$(H$_2) \ge 18$. In other words, DLAs with H$_2$ column densities below this value can be considered as non-H$_2$-bearing. The observed composite spectrum is therefore described as
\begin{equation}\label{rate}
S'(\lambda) = 1 - r \cdot (1 - S(\lambda)).
\end{equation}
We remark that the effect of non-unity incidence rate $r$ is similar to that of partial coverage \citep[e.g.][]{Balashev2011}: a fraction of the background light is not absorbed at all at the position of H$_2$ lines, acting like a shifted zero-level (see Fig.~\ref{SDSS}).
While H$_2$ absorption systems include lines from different H$_2$ rotational levels, 
most of the total H$_2$ is found in the $J=0$ and $1$ levels, which are saturated. Higher rotational levels are therefore included for illustrative purposes (assuming typical observed excitation) but not used to constrain the fit. The relative population of the $J=0$ and $J=1$ levels is held fixed to 0.3~dex, which is typical for high-$z$ DLAs with $\log N($H$_2)>18$ and corresponds to the kinetic temperature in the cold ISM ($T \sim 100$\,K). The Doppler parameter is fixed to $b=4$\,km/s, corresponding to typically observed values for H$_2$. 
We are therefore left with only two parameters that determine the synthetic composite profile for H$_2$: the power law slope, $\beta$, and the incidence rate, $r$. We fit the H$_2$ signal using 
the above formalism and obtain constraints on $r$ and $\beta$ parameters (see 
Fig.~\ref{SDSS}). 
We note that we used a simplified model, and the fact that we kept some parameters fixed
introduces systematic uncertainty. 
In the following, we conservatively estimate the systematic uncertainty by varying $\log N_{\rm up}$ and $\log N_{\rm low}$ by 0.5 dex and the effective Doppler parameter between 2 and 8\,km\,s$^{-1}$. 

\section{Discussion \label{discussion}}

\subsection{Incidence rate of molecular hydrogen}
\label{inc_rate}
We find that the incidence rate of H$_2$ is $r\sim 4.0\pm0.5(\rm stat)\pm1.0(\rm sys)\,\%$, $\sim 8.2\pm0.7(\rm stat)\pm1.5(\rm sys)\,\%$ and $\sim 37\pm5(\rm stat)\pm4(\rm sys)\,\%$ for the {\sl DLAs}, {\sl metal-DLAs} and {\sl ESDLAs} samples, respectively (with statistical uncertainties corresponding to 0.683 confidence level).

First of all, taken at face value, the overall incidence rate appears smaller than the observed detection rate for DLAs observed at high-spectral resolution with UVES 
\citep[$\sim 10\%$,][]{Ledoux2003,Noterdaeme2008}. This could result from the different $N(\HI)$-distributions: the composite spectrum includes a large fraction of sub-DLAs while the UVES sample is skewed towards higher $N(\HI)$. 
In addition, the UVES sample is relatively small and heterogeneous and includes systems with $\log N($H$_2)<18$ as well. Counting only for $\log N($H$_2)\ge 18$ systems, and assuming the same simple correction of $N(\HI)$ distribution as in \citet{Noterdaeme2008}, the incidence rate of strong H$_2$ systems in the UVES sample is around $7\pm 4$\%. 

Secondly, there is only one confident H$_2$ detection in the search for H$_2$ in 86 SDSS DLAs by \citet{Jorgenson2014}. 
However, a large fraction of their observations (but the detection) were done at medium spectral resolution with MagE on the Magellan telescope and the detection limits are spread over a range which may not be good enough to safely take the observed detection rate as value for the true incidence rate at $\log N$(H$_2)\sim 18$, which the authors argue to be $<6\%$ at 95\% confidence level, actually consistent with our measurement. 

Directly searching for damped H$_2$ absorption in the SDSS spectra of DLAs, \citealt{Balashev2014} derive a H$_2$ detection rate $\sim9\%$ at $\log N($H$_2) \sim  18.5$ ($7\%$ at $\log N($H$_2) \sim 19$). However, as they caution, quality of SDSS spectra leads to that the automatic procedure overestimates $N(\rm H_2)$, meaning that the derived values rather represent upper limits on the actual incidence rates and hence consistent with our current result. 
Finally, our measurement based on the DLA composite spectrum may be biased towards low detection rate because of the presence of false positive DLAs contributing to the stack (or equivalently systems that actually have $N(\HI)<10^{20}$cm$^{-2}$).  

A careful assessment of the biases or selection function of the different works is therefore necessary to derive the true incidence rate, in particular since this value is sensitive to the low $N(\HI)$ systems that represent the main contribution to the DLA counting statistics.
Irrespective of the absolute incidence rate, the relative increase in the detection rate for systems with prominent metal lines is at least in qualitative agreement with the finding by \citet{Petitjean2006} that H$_2$ is more frequently seen in high-metallicity DLAs in which higher  
dust abundance is expected \citep[e.g.][]{Ledoux2003}. Indeed, dust grains act as a catalyst for the formation of H$_2$ and provide efficient shielding from UV photons.  
We also find an order of magnitude increase in the incidence rate for ESDLAs compared to the DLA sample. 
We confirm this both through visual inspection of individual ESDLA spectra and our automatic search for H$_2$ \citep[see][]{Balashev2014}. 
The strong increase of the H$_2$ incidence rate at high \HI\ column densities is qualitatively consistent with steady-state models for \HI-H$_2$ conversion, discussed in Sect.~\ref{HIH2}. 

\subsection{The H$_2$ distribution function}
From the incidence rate and slope of the power-law distribution, $\beta = -1.29 \pm 0.06(\rm stat) \pm 0.10 (\rm sys)$\footnote{The slopes obtained for the {\sl metal-DLAs} and {\sl ESDLAs} samples agree with this value within statistical and systematic uncertainties.}, we can estimate the H$_2$ column density distribution function $f_{\rm H_2}(N)$ (the number of H$_2$ absorbers per unit column density per unit absorption distance) in the range $\log N_{\rm low}-\log N_{\rm up} = 18-22$. 
The normalisation can be obtained by matching $f_{\rm H_2}(N)$ with the observed \HI\ column density distribution function $f_{\rm HI}(N)$ of the DLA sample that was used to construct the composite spectrum as
\begin{equation}\label{r_def}
\int\limits_{10^{18}}^{10^{22}}  f_{\rm H_2}(N) dN \equiv \frac{d\rm \cal{N}_{\rm H_2}}{dX} = r \cdot \frac{d\rm \cal{N}_{\rm HI}}{dX} \equiv r \cdot \int\limits_{10^{20}}^{10^{22}} f_{\rm HI}(N) dN,
\end{equation}
where $\cal{N}$ is the number of absorbers, $X$ is the absorption distance (see e.g. Eq.~2 of \citealt{Noterdaeme2009}) and  $f_{\rm HI}(N)$ is taken from \citet{Noterdaeme2012}, with the high $N(\HI)$-end updated by \citet{Noterdaeme2014}. 

\begin{figure}
\centering
\includegraphics[trim={0.5cm 0.7cm 0.4cm 0.6cm},clip,width=\hsize]{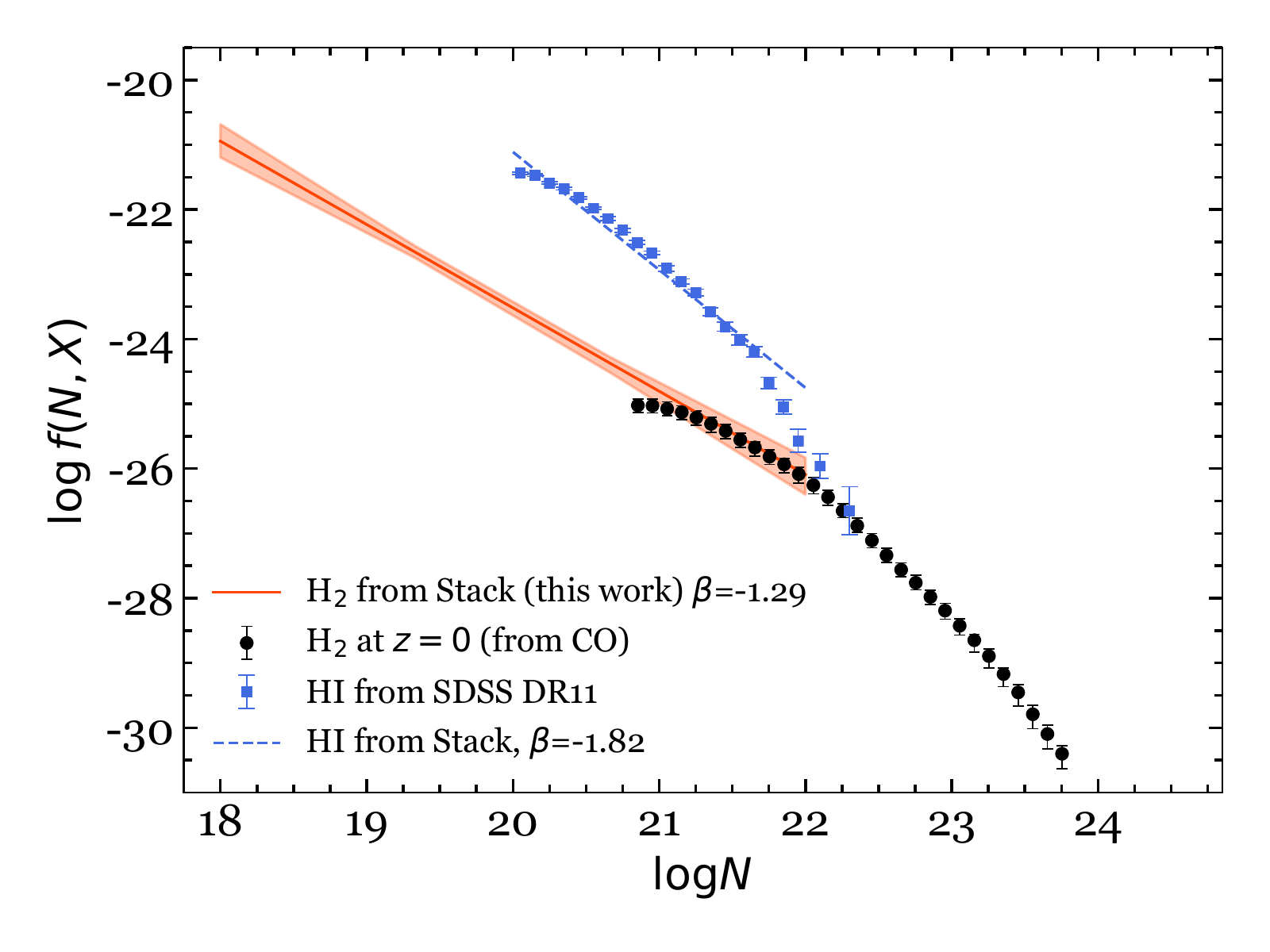}
		\caption{Column density distribution functions of H$_2$ and H\,{\sc i}. The blue points correspond to the high-$z$ \HI\ distribution function from DR9 \citet{Noterdaeme2012} with the high $N(\HI)$-end updated using DR11 by \citet{Noterdaeme2014}. The blue dashed line is our current derivation using the single fit of the composite Lyman-$\alpha$ absorption from \citet{Mas-Ribas2017}, see Fig.~\ref{HIstack}. The black dots correspond to the H$_2$ distribution function derived at $z=0$ from CO maps \citet{Zwaan2006}. Finally, the red line with shaded uncertainties correspond to our measurement of the H$_2$ distribution function at high-$z$.
		\label{fNH}}
\end{figure}

Fig.~\ref{fNH} shows our derived $f_{\rm H_2}(N)$ at $z\sim 3$ from the {\sl DLA}-sample, 
along with the $f_{\rm H_2}(N)$ derived by \citet{Zwaan2006} at $z=0$ for $\log N($H$_2)>22$ using CO emission maps of local galaxies, corrected using the updated value of $X_{\rm CO}$ factor given by \citet{Bolatto2013}. 
The \HI\ distribution function and our derivation using a simple power-law fit of 
the composite Ly-$\alpha$ line are also shown. 
While our method does not constrain the H$_2$ distribution function at $\log N($H$_2) > 22$, 
one can see that our derived $f_{\rm H_2}(N)$ connects with that at $z=0$ from CO emission\footnote{While not providing much details about their derivation of $f_{\rm H_2}(N)$ at $\log N (\rm H_2) \sim 18$, it seems that for this range \citet{Zwaan2006} actually derived the H\,{\sc i} column density distribution of H$_2$-bearing DLAs (i.e. $f_{\rm H\,{\sc i}}(N)$) instead, which is 2-3 orders of magnitude below the actual $f_{\rm H_2}(N)$.}.
This is remarkable and suggests no or little evolution in the comoving number density of H$_2$ systems, at least in the range $\log N$(H$_2) \sim 21-22$. This could indicate a self-regulating mechanism but may actually be only coincidental due to several factors acting in opposite directions: on the one hand, the H$_2$ abundance in the diffuse ISM at high-$z$ can be lower compared to $z=0$ due to the lower metallicities and higher UV background, however, on the other hand, the total incidence of neutral gas is a few times higher at high-$z$. In addition, there are likely observation biases against high column densities. 
Both line absorption by damped H$_2$ and \HI\ lines as well as continuum extinction by the associated dust can result in the exclusion of line of sights with high $N($H$_2$) from photometrically-selected surveys (as already discussed in several papers, e.g. \citealt{Pei1991, Pontzen2009, Noterdaeme2015b}).

\subsection{$\Omega_{\rm H_2}$}
Ideally, one would like to derive the cosmological density of molecular hydrogen, $\Omega_{\rm H_2}$, by integrating $N \cdot f_{\rm H_2}(N)$ over all column densities. However, $\Omega_{\rm H_2}$ does not converge by extrapolating the power-law slope to higher column densities, since $\beta$ is shallower than -2. This means that $f_{\rm H_2}(N)$ must steepen at a column density higher than what is probed in this work. 
Assuming that the steepening point of $f_{\rm H_2}(N)$ at high-$z$ is at the same column density as at $z=0$ ($\log N\sim 23$), we estimate the fraction of $\Omega_{\rm H_2}$ in the "diffuse" molecular gas ($\log N($H$_2)<22$) to be $\sim15\%$, i.e. most of the H$_2$ should reside in very high column density systems that still escape direct detection through UV lines. 

\subsection{The H\,{\sc i}/H$_2$ transition}
\label{HIH2}
The order-of-magnitude higher incidence rate of H$_2$ for ESDLAs suggests that these column densities are close to the \HI-H$_2$ transition point for typical conditions\footnote{The transition can be observed at much lower $N$(\HI) in systems with very high metallicity, as in \citet{Noterdaeme2017}.} in the neutral gas at high redshifts.  A statistical estimate of the transition point can also be obtained by extrapolating the derived power law for $f_{\rm H_2}(N)$ which should equate $f_{\rm H\,{\sc i}}(N)$ at around $\log N($H\,{\sc i},H$_2) \sim 22.3$. In other words, the derived H\,{\sc i} and ${\rm H_2}$ distribution functions suggest that there are more DLAs with $\log N(\rm H_2)>22.3$ than with $\log N(\rm HI)>22.3$, statistically meaning that, for a given system with $\log N(\rm H)>22.3$, H$_2$ will be dominant. This value is consistent with the expected critical surface density where \HI\ converts into H$_2$, $\Sigma_{\rm HI}^{crit} \sim 10/Z$~M$_{\rm \odot}$\,pc$^2$ \citep{McKee2010}, for the typical metallicity for DLAs \citep[$Z \sim 1/20$,][]{Rafelski2012, DeCia2018}. 
The fraction of ESDLAs reaching such high column densities is very small, only a couple of such systems appear in our ESDLA sample, explaining why the \HI/H$_2$ transition remains mostly elusive in neutral gas with typical low-metallicities at high redshifts. 
 However, selection biases in the SDSS can also play a significant role on the derived incidence rate at such column densities, impeding a more robust quantitative estimate of the characteristic transition column density. 

\section{Concluding remarks}

We have quantified, for the first time, the column density distribution of H$_2$ in the range $\log N(\rm H_2) = 18 - 22$ at $z\sim 3$ directly from composite spectra built from a large homogeneous sample of DLAs from the SDSS without the need of detecting individual H$_2$ systems. 
However, we caution that our study inevitably inherits the biases from the selection function of its parent quasar sample. This likely affects mostly 
the high $N($H$_2$) end of the $f_{\rm H_2}(N)$ distribution function that contains the bulk of the molecular gas.  
The strong dimming of the background sources by foreground molecular gas will indeed make it very difficult to directly observe in UV the column densities higher than those probed in this work. 
Such systems can in principle be detected in absorption in the sub-millimeter using other molecular tracers, such as the well-known case at $z=0.89$ towards  PKS\,1830$-$211 \citep{Wiklind1998}, 
but the small cross-section of these systems together with the paucity of strong continuum sources makes this a very rare event.
Observationally and blindly determining $\Omega_{\rm H_2}$ therefore constitutes a very challenging task for astronomers.

\section*{Acknowledgements}
We thank Lluis Mas-Ribas for sharing his DLA composite spectra. We thank the referee for insightful comments that helped improving the clarity of this paper
and \mbox{J.-K.} Krogager for useful comments and discussions. 
This research is supported by the French {\sl Agence Nationale de la Recherche}, under grant ANR-17-CE31-0011-01 (project ``HIH2''). SB is supported by RFBR (grant No. 18-02-00596). PN is grateful to the {\sl Programme National de Cosmologie et Galaxies}, funded by CNRS/INSU-IN2P3-INP, CEA and CNES, France for support and to the Ioffe Institute for hospitality during the time part of this research was done.



\bibliographystyle{mnras}

\bibliography{library} 
\bsp	
\label{lastpage}
\end{document}